\documentclass{rnaastex}

\begin{document}

\title{Multicolor Photometry of the Neptune Irregular Satellite Neso\footnote{In press in RNAAS, accepted for publication March 20${\mathrm{th}}$, 2018.}
}

\correspondingauthor{Michele Maris}
\email{michele.maris@inaf.it}

\author{Michele Maris}
\affiliation{INAF- Osservatorio Astronomico di Trieste, Via G.B.Tiepolo 11, I34100 Trieste, Italy}

\author{Giovanni Carraro}
\affiliation{Dipartimento di Fisica e Astronomia, Università degli Studi di Padova, Italy}

\author{Mario Melita}
\affiliation{Instituto de Astronomía y Física del Espacio (CONICET-UBA), CABA, 1428, Buenos Aires, Argentina}

\author{Gabriela Parisi}
\affiliation{Instituto Argentino de Radioastronomía (CCT-La Plata, CONICET), C.C. No. 5, 1894 Villa Elisa, Prov. de Buenos Aires, Argentina}


\section{}

\begin{figure}[h!]
\begin{center}
\includegraphics[width=180mm,angle=0]{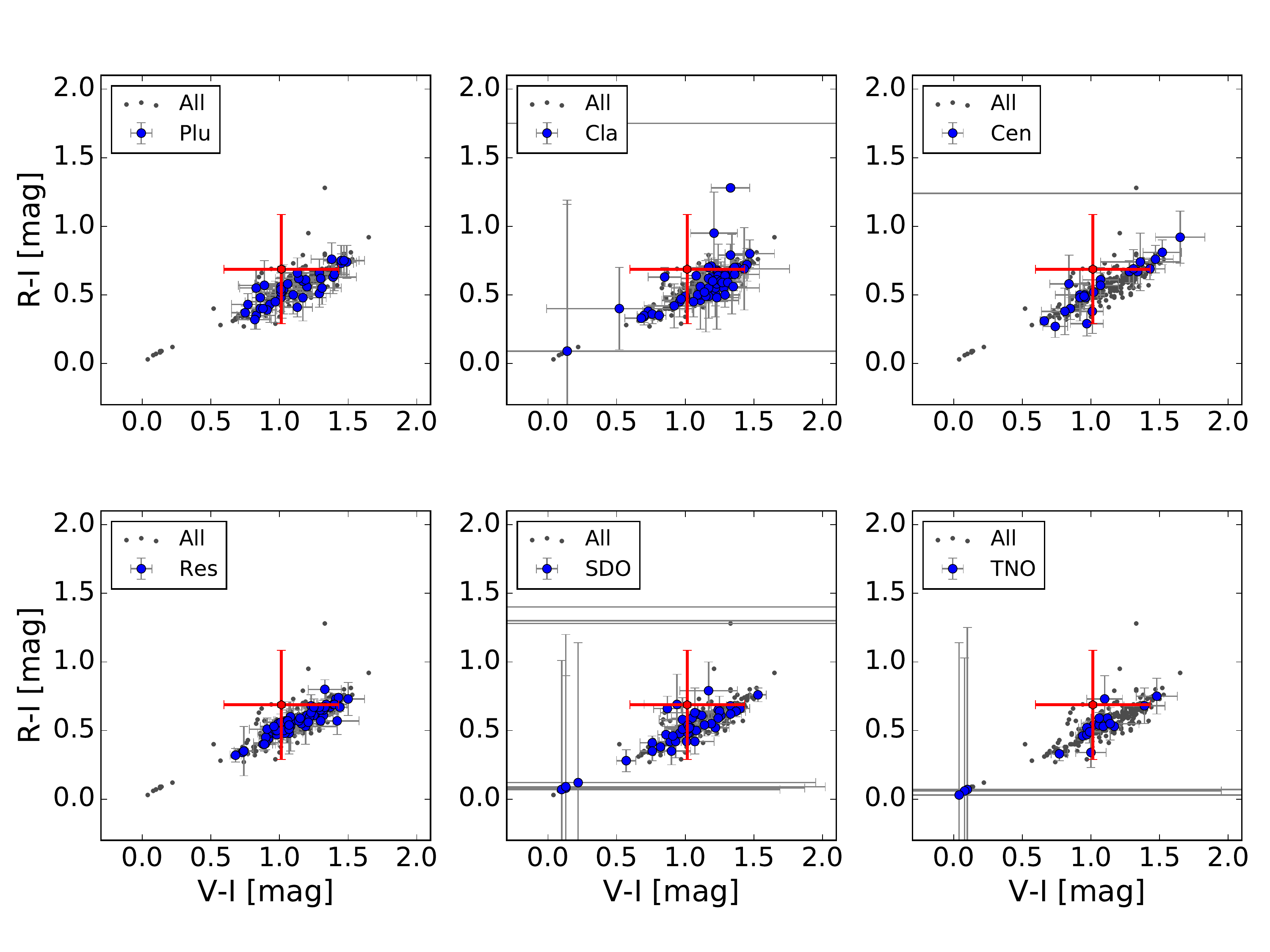}
\caption{\label{fig:1}
A comparison of the Neso colors from this work with solar system minor body
classes from \citet{2015A&A...577A..35P}.
In each frame Neso colors are marked in red and are compared with the colors
of a single class
from \citet{2015A&A...577A..35P} represented by blue points, the gray points
represents the other classes and are left as a reference.
From left to right and from top to bottom: 
Plutinos (Plu),
Classical KBOs (Cla),
Centaurs (Cen),
Resonant (Res),
Scattered Disk Objects (SDO),
Trans Neptuninan Objects (TNO). 
All the VRI photometry used to derived the average colors of Neso shown in red are listed in Tab.~\ref{tab:1}
}
\end{center}
\end{figure}

We report on time series photometry of the faint Neptune irregular satellite Neso. 
Observations in the V, R, and I pass-bands were performed in photometric conditions at the Cerro Paranal observatory using the instrument FORS2. Photometric calibration was secured using observatory standards.  Magnitudes were extracted using the IRAF task {\it qphot}, and corrected for aperture using bright stars in the field. Photometric errors have been derived extracting photometry of stars with magnitudes comparable to Neso, and were found to be around 0.3$-$0.4 mag.
Astrometry is derived using up to 5 stars in the satellite vicinity basing on the astrometric catalog USNO-B1. 
We expect these data will be useful to better constrain the poorly known Neso orbit \citep{2017DDA....4820504B}.
The time coverage is not sufficient to construct a light curve and derive a meaningful rotational period.
However, we could derive  new estimates of apparent magnitudes, in particular in V and I passbands,  and for the first time we could calculate
Neso colors. 
Arithmetic averages yields  $V=25.6\pm0.3$ from 12 exposures, $R=25.2\pm0.2$ from 13 exposures,
and  $I=24.5\pm0.3$ from from 25 exposures. 
The R averaged magnitude is in agreement with \citet{2011AJ....141..135B} who
quote R=25.2, the only available measure to date.
From these  measures of the averaged magnitudes, we derive averaged colors.
We obtain $V-I=1.0\pm0.4$, $R-I=0.7\pm0.4$, and $V-R=0.3\pm0.4$, respectively.
The color $R-I$ appear to be slightly redder
than the typical values  for Centaurs and KBOs as reported in  \citet{2015A&A...577A..35P}, while the color $V-I$ is in nice agreement
with both populations (see Fig.~\ref{fig:1}).
Given the large error-bars in the averaged colors, it is a difficult task to assign Neso to any of the \citet{2015A&A...577A..35P} classes, although the data seem to suggest that we can rule out its membership in classes of resonant objects or Plutinos.


\startlongtable
\begin{deluxetable}{ccccccc}
\tablecaption{Astrometric and photomeric data for NESO, 2010 July 15\label{tab:1}}
\tablehead{
\colhead{UT} & \colhead{RA(J2000.0)} & \colhead{Dec(J2000.0)} & \colhead{Filter} & \colhead{Airmass} & \colhead{Exp.Time (secs)} & \colhead{Mag}
}
\startdata
02:58:04.455 &  22:01:37.34  &   -12:26:34.87  &    R  &  1.920  &  600  &  25.4 \\
03:08:40.521 &  22:01:37.26  &   -12:26:35.03  &    I  &  1.798  &  600  &  24.1 \\
03:20:20.734 &  22:01:37.33  &   -12:26:34.51  &    V  &  1.683  &  600  &  25.2 \\
03:31:03.481 &  22:01:37.19  &   -12:26:35.58  &    R  &  1.593  &  600  &  25.3 \\
03:41:39.538 &  22:01:37.17  &   -12:26:35.82  &    I  &  1.551  &  300  &  24.8 \\
03:47:05.577 &  22:01:37.14  &   -12:26:35.86  &    I  &  1.513  &  300  &  24.6 \\
03:53:43.412 &  22:01:37.12  &   -12:26:35.96  &    V  &  1.439  &  600  &  25.5 \\
04:04:26.449 &  22:01:37.08  &   -12:26:36.12  &    R  &  1.380  &  600  &  25.2 \\
04:15:02.517 &  22:01:37.06  &   -12:26:36.27  &    I  &  1.352  &  300  &  25.1 \\
04:20:30.705 &  22:01:37.03  &   -12:26:36.40  &    I  &  1.327  &  300  &  25.1 \\
04:27:36.114 &  22:01:36.99  &   -12:26:36.69  &    V  &  1.276  &  600  &  25.7 \\
04:38:18.511 &  22:01:36.95  &   -12:26:36.74  &    R  &  1.236  &  600  &  25.1 \\
04:48:54.437 &  22:01:36.93  &   -12:26:36.93  &    I  &  1.217  &  300  &  24.3 \\
04:54:23.057 &  22:01:36.91  &   -12:26:37.14  &    I  &  1.200  &  300  &  24.4 \\
05:01:18.504 &  22:01:36.84  &   -12:26:40.47  &    V  &  1.166  &  600  &  25.4 \\
05:12:01.492 &  22:01:36.82  &   -12:26:37.47  &    R  &  1.139  &  600  &  24.7 \\
05:22:37.448 &  22:01:36.79  &   -12:26:37.52  &    I  &  1.127  &  300  &  24.3 \\
05:28:05.997 &  22:01:36.78  &   -12:26:37.51  &    I  &  1.115  &  300  &  24.2 \\
05:35:21.816 &  22:01:36.76  &   -12:26:37.61  &    V  &  1.093  &  600  &  25.1 \\
05:46:04.414 &  22:01:36.71  &   -12:26:38.14  &    R  &  1.076  &  300  &  25.0 \\
05:56:40.521 &  22:01:36.69  &   -12:26:38.28  &    I  &  1.068  &  300  &  24.8 \\
06:02:08.589 &  22:01:36.67  &   -12:26:38.12  &    I  &  1.061  &  300  &  24.9 \\
06:08:54.996 &  22:01:36.62  &   -12:26:38.75  &    V  &  1.048  &  600  &  25.3 \\
06:19:37.303 &  22:01:36.59  &   -12:26:38.96  &    R  &  1.039  &  600  &  25.4 \\
06:30:13.530 &  22:01:36.56  &   -12:26:39.23  &    I  &  1.035  &  300  &  24.2 \\
06:35:41.979 &  22:01:36.54  &   -12:26:39.08  &    I  &  1.032  &  300  &  24.0 \\
06:42:27.105 &  22:01:36.52  &   -12:26:39.25  &    V  &  1.026  &  600  &  25.8 \\
06:53:09.383 &  22:01:36.46  &   -12:26:39.55  &    R  &  1.024  &  600  &  25.6 \\
07:03:45.499 &  22:01:36.44  &   -12:26:39.77  &    I  &  1.023  &  300  &  23.8 \\
07:09:13.988 &  22:01:36.42  &   -12:26:39.74  &    I  &  1.023  &  300  &  24.7 \\
07:15:47.623 &  22:01:36.39  &   -12:26:40.03  &    V  &  1.025  &  600  &  25.5 \\
07:26:30.371 &  22:01:36.36  &   -12:26:40.13  &    R  &  1.029  &  600  &  25.3 \\
07:37:06.497 &  22:01:36.32  &   -12:26:40.43  &    I  &  1.031  &  300  &  24.5 \\
07:42:34.687 &  22:01:36.30  &   -12:26:40.45  &    I  &  1.035  &  300  &  24.5 \\
07:49:14.322 &  22:01:36.28  &   -12:26:40.39  &    V  &  1.043  &  600  &  25.6 \\
07:59:57.409 &  22:01:36.29  &   -12:26:40.59  &    R  &  1.054  &  600  &  25.1 \\
08:10:33.617 &  22:01:36.21  &   -12:26:40.92  &    I  &  1.060  &  300  &  24.5 \\
08:16:01.956 &  22:01:36.18  &   -12:26:41.10  &    I  &  1.067  &  300  &  24.5 \\
08:22:41.182 &  22:01:36.15  &   -12:26:41.40  &    V  &  1.084  &  600  &  25.8 \\
08:33:23.428 &  22:01:36.11  &   -12:26:41.45  &    R  &  1.102  &  600  &  25.1 \\
08:43:59.605 &  22:01:36.10  &   -12:26:41.54  &    I  &  1.113  &  300  &  24.5 \\
08:49:28.015 &  22:01:36.06  &   -12:26:41.43  &    I  &  1.124  &  300  &  24.8 \\
08:56:06.490 &  22:01:36.03  &   -12:26:41.81  &    V  &  1.151  &  600  &  25.8 \\
09:06:49.418 &  22:01:35.99  &   -12:26:42.20  &    R  &  1.179  &  600  &  25.3 \\
09:17:25.754 &  22:01:35.97  &   -12:26:42.14  &    I  &  1.196  &  300  &  24.9 \\
09:22:54.244 &  22:01:35.94  &   -12:26:42.55  &    I  &  1.213  &  300  &  24.7 \\
09:29:31.409 &  22:01:35.92  &   -12:26:42.67  &    V  &  1.252  &  600  &  25.9 \\
09:40:14.527 &  22:01:35.87  &   -12:26:42.90  &    R  &  1.294  &  600  &  25.4 \\
09:50:50.793 &  22:01:35.84  &   -12:26:42.78  &    I  &  1.318  &  600  &  24.6 \\
09:56:18.652 &  22:01:35.81  &   -12:26:43.17  &    I  &  1.343  &  600  &  24.6 \\
\enddata
\end{deluxetable}

\acknowledgments
Based on observations carried out at ESO Paranal under program 085.C-0187(C).

\end{document}